# Implementation of Ethically Aligned Design with Ethical User stories in SMART terminal Digitalization project: Use case Passenger Flow

Erika Halme,
Faculty of Information Technology, University of Jyväskylä, Jyväskylä, Finland
0000-0003-0750-1580

Jani Antikainen,
Faculty of Information Technology, University of Jyväskylä, Jyväskylä, Finland
0000-0003-3367-0492

Kai-Kristian Kemell,
Faculty of Information Technology, University of Jyväskylä, Jyväskylä, Finland
0000-0002-0225-4560

Mamia Agbese,
Faculty of Information Technology, University of Jyväskylä, Jyväskylä, Finland
0000-0002-5479-7153

Marianna Jantunen,
Faculty of Information Technology, University of Jyväskylä, Jyväskylä, Finland
0000-0002-8991-150X

Ville Vakkuri,
Faculty of Information Technology, University of Jyväskylä, Jyväskylä, Finland
0000-0002-1550-1110

Hanna-Kaisa Alanen,
Faculty of Information Technology, University of Jyväskylä, Jyväskylä, Finland
0000-0002-8797-3434

Arif Ali Khan,
Faculty of Information Technology, University of Jyväskylä, Jyväskylä, Finland
0000-0002-8479-1481

Pekka Abrahamsson,
Faculty of Information Technology, University of Jyväskylä, Jyväskylä, Finland
0000-0002-4360-2226

## Abstract

Digitalization and Smart systems are part of our everyday lives today. So far the development has been rapid and all the implications that comes after the deployment hasn't been able to foresee or even assess during the development, especially when ethics or trustworthiness is concerned. Artificial Intelligence (AI) and Autonomous Systems (AS) are the direction that software systems are taking today. It is witnessed in banks, stores, internet and it is proceeding to transportation as well as on traveling. Autonomous maritime industry has also taking this direction when taking under development in digitalization on fairway and port terminals. AI ethics has advanced  profoundly since the machine learning develop during the last decade and is now being implemented in AI development and workflow of software engineers. It is not an easy task and tools are needed to make the ethical assessment easier. This paper will review a research in an industrial setting, where Ethically Aligned Design practice, Ethical User Stories are used to transfer ethical requirements to ethical user stories to form practical solutions for project use. This project is in the field of maritime industry and concentrates on digitalization of port terminals and this particular paper focuses on the passenger flow. Results are positive towards the practice of Ethical User Stories, drawn from a large empirical data set.

**Keywords:** Digitalization, Artificial Intelligence, SMART systems, Ethics, User Stories, Passenger Flow, Port terminal

# 1. Introduction

Digitalization and especially Artificial Intelligence (AI) is today been undertaking in many project configurations and the benefits that the AI has declared to bring on the society, industry or for the environment has been taken as a way to reach goals in many digitalization involved development projects. One area of interest for digitalization and SMART systems is port terminal and it's different components. Digitalization of traditional port terminal to a future SMART Terminal has now been under development and witnessed in a project called the SMARTER. This project is a research and design (R&D) influenced project, which is a part of a larger program, Sea for Value program (DIMECC Oy, 2021), both aiming for digitalization and increased level of autonomy (DIMECC Oy, 2021), where "The mission of SMARTER is to create replicable models for digitalization" for future terminals [1]. Objectives on the program is to develop "safer, more efficient, sustainable, and reliable service chains to meet the requirements for a better quality of life and global prosperity". (DIMECC Oy, 2021)

So today we are aiming for a better, prosperous and safe society, which is often approach and pursued through technological innovation, but e.g. AI, also called as autonomous systems (AS), has been advancing since the 1950's the impact assessment on using these systems as well as taking the assessment process to the development of AI/AS system has only in recent years been under interest of various institutes, organizations and governing bodies producing different sets of guidelines, principles and even some regulations. (Jobin, 2019) Implementing these principles etc. to the workflow of e.g. software engineers has been the main research focus in the University of Jyväskylä's (JYU) AI Ethics lab since 2018 following several publications and tools to provide for the practitioners in the field as well as for the academics sharing the same interest.

AI ethics is a renewed branch of research area in the field of information technology. It is multidisciplinary research area that shares views from information ethics, Computer Sciences, AI research, Software development, even human rights. AI ethics can be approached from several points of view and when designing these artefacts the Dignum paradigm (Dignum, 2018) presents one way of consideration. There, ethics and AI are related in three levels as in Ethics by Design, Ethics in Design and Ethics for Design, where we in JYU place ourselves in the middle ground of Ethics in and for Design, as we develop tools to transform ethical principles to practical and concrete means in Ethically Aligned Design (EAD) in AI.

One of the tools that has risen from our studies is Ethical User stories (Halme et al., 2021). Ethical User Stories relates to Agile requirements engineering, where requirements for Software Development are processed through a User story process using a framework of ethical principles that are connected to technology design or development. AI ethics Lab in JYU has described in their article by Halme et al. "How to Write Ethical User Stories: Impacts of the ECCOLA method" that: "Writing user stories is a practice commonly used to help define requirements during development, especially in Agile software development. Thus, we felt that ethical user stories could be one way of making (AI) ethics a part of the workflow of developers." This Ethical User story practice was implemented in an industrial setting in this article in the previously mentioned SMARTER project. The following chapters will describe more closely the practice background and premises.

The project in question produces and shares information between ten (10) industrial partners and five (5) research institutes with the lead of R&D company. University of Jyväskylä (JYU) was responsible for the Ethically Aligned Design, where the main target of the study was to introduce the requirements for ethically aligned design in Smart terminals. According to (DIMECC Oy, 2021) Smarter project aims for reduction on emissions and to exceptional flow and experience for the passengers and cargo by providing concrete research based recommendations on business, data usage & sharing. Our AI ethics lab research group at JYU has approached this mission and concrete recommendations through Software development methods, more precisely Agile Software development. As mentioned in the previous chapter, in practice we approach this responsibility via Ethical User Story Process, where the empirical material was collected by the project industrial and institutional partners in a virtual workshop. This is described in detail in figure 1. Through the practice of Ethical User Stories (EUS) the project will receive tangible outcome of the EAD study, the research based recommendations.

These Ethical User Stories were initiated by a project workshop notes, the raw data, that was the first step in the workflow towards responsible and Ethically aligned SMART terminal development. This ethical user story process can be considered ad hoc and developed as the project evolves but follows the utilized AI ethics principles powered design tool, ECCOLA. This tool is developed by the AI Ethics lab in JYU and used in several AI and SE projects since it's first version was introduced to the public in 2018. In practice, ECCOLA is a method that uses gamifieng technique that empowers ethical thinking in the product development process. It has 21 cards to choose from with a content of AI ethics principles, 8 themes with each theme holding 1-6 card topics, also AI ethics principles. According to ECCOLA Game Sheet the method is intended to be used during the entire design and development process in three steps: 1. prepare, as choosing the relevant cards, 2. review, as keeping the chosen cards in hand during single task, and 3. evaluate, as reviewing to check if all planned actions are taken. (Vakkuri et al., 2021) In relation to Ethically Aligned Design, Ethical user story practice can actually use any ethics related design/development framework from the field of engineering or technology design, but as ECCOLA is developed particularly to AI software development, and in the perspective of agile software development, for us it was natural to use this method in the process.

This article is then motivated by the program objectives and the growing trend of AI enhanced smart systems, where the impacts of utilization of these systems is requiring concrete actions from research towards ethically aligned design and trustworthy smart system environment. Inspired by the field's will for ethically aligned design towards AI/AS systems development and by the practitioners' need for tools to conduct their work towards ethically aligned design the main research questions were unfold to: How to write ethical user stories? and What are the ethical requirements in Maritime industry, especially in port terminals when switching over to SMART terminals?

This is the first article published in the series of three similarly structured and content wise articles and based on the SMARTER project's division of development work. This article reports and shares the outcome of a passenger flow use case study in terms of ethical requirements towards the SMARTER project research. The project and the division of work is described more in detail in section 3.

The article is organized as follows: section one introduce the research motivation and main goal for the study. Section two provides the theoretical background for the research, while section three goes through the research framework and study setting. Section four then elaborates the results and section five concludes the article with discussion on the research outcome.

## 2. Theoretical Background

Passenger Flow

As the number of people using transport systems increases, the flow of passengers is critical to the success of an effective transport system as it affects movement through different service points. Passenger flow (PF) involves navigating passengers (people, vehicles, cargo, ships, machinery, and all the associated operations); and consists of the experience at each touchpoint in the process. PF is becoming vital for resource scheduling, planning, public safety, and risk management in transport systems (Du et al., 2019). Some of its critical application areas include prediction and forecasting to enable transport systems to significantly manage their operations and resources (Li et al., 2018). One of the main applications of PF is as a prediction tool in urban transport, where it forms part of an essential technological resource for enabling sustainable and steady development in transportation (Lin et al., 2020). As a prediction tool, PF tends to study patterns of crowd travel behavior and form real-time traffic operation state evaluation, which can contribute to schedule resources in traffic management (Li et al., 2018).

Passenger flow can be affected by various dynamic and complex factors, including dynamic traffic routes, upgrade of transportation facilities, multifaceted transfer flows, rush hour, and external influences such as bad weather (Du et al., 2019). However, dynamic changes in global transport infrastructure alongside the widespread adoption of digital technologies are increasingly influencing how passenger flow is designed (Li et al., 2018). The continuous growth of Industry 4.0 coupled with unprecedented challenges such as COVID-19 means that the traditional methods of designing passenger flow systems can no longer suffice and need to be replaced or complemented with digital options in new technological spaces (Krile et al. 2021). Part of this innovative way of designing PF systems involves implementing intelligent technologies to help manage operations and maintain facilities safely, securely, and environmentally friendly (Molavi et al., 2020).

This change in design approach from statistics to technologies such as Artificial Intelligence (AI) has resulted in most studies in the review of the literature to focus on Machine Learning (ML) and Deep Learning (DL) techniques to enhance PF ( Liu et al., 2019, Li et al., 2018, Cheng et al. 2021, Xie et al. 2020). While some studies have explored simulation approaches (Rexfelt,

2014), most have embraced AI and associated technologies in designing PF systems. As a result, PF systems are becoming an interconnected smart environment involving big data, the Internet of Things (IoT), and AI that facilitates autonomous practices (de Barcelona, 2012).

However, designing AI-enabled passenger flow systems can be challenging due to the ethical concerns connected with AI and associated technologies, such as ML. ML practices operate mindlessly with no conscious understanding of the broader context of their processes and cannot contemplate their actions' ethics (Asatiani et al., 2021). In addition, the opaque nature of ML-powered AI systems may create a challenge in apportioning responsibilities or explanations to relevant actors in a case of redress or failure in sensitive situations resulting in diminished customer confidence. Privacy and bias represent critical ethical challenges that developers tasked with designing AI-enabled PF systems may encounter (Clever et al., 2018). Privacy and trust issues may emanate from unethical customer data collection, violating users' privacy (Culnan, 2019). Users of the system may question the various surveillance systems such as videos and cameras engaged in collecting passenger flow data as invasive and their application. In addition, AI reportedly has a history of unfairness regarding ethnicity, gender, and race (Sen, Dasgupta & Guptal, 2020) which may spark ethical concerns about the quality of data and the bias that may ensue from its application. The EU analyses that sustained use of AI systems could lead to breaches of fundamental human rights if ethical issues remain unrestricted (European Commission, 2020).

## Ethically Aligned Design

With the increasing calls for ethically aligned designs, several organizations, governments, and research bodies have attempted to tackle ethical AI concerns. One of the results is the principles approach, where principles serve as guidelines in developing and deploying AI systems (cite). However, developers still struggle to transition principles to practice due to a lack of actionable tools and methods (Vakkuri, 2021). Furthermore, there is also a risk for unethical AI behavior at the design stage, raising the issue of which ethical principle should be considered pertinent in designing and managing AI-based systems (Brendel et al., 2021). As a possible solution to address these concerns, the ECCOLA method has been developed to action principles to practice as a tool that aids the ethically aligned design of AI systems (Vakkuri et al., 2021). ECCOLA was created by distilling fundamental AI ethics principles to guidelines for AI developers at the design and development stages (Vakkuri et al., 2021). In addition, ECCOLA can be instrumental in producing (ethical) user stories to help define high-level requirements and, in the process, translate ethical principles to tangible design requirements for developers of AI systems (Halme et al., 2021).

In addressing the ethical challenges associated with designing AI-enabled PF systems, the ECCOLA method can serve as a solution for translating high-level requirements to system requirements for developers in the form of user stories (Halme et al., 2021). ECCOLA as a user story tool can prove instrumental in facilitating communication in defining high-level requirements for AI systems developers. This practice can also serve as an ethical guide to help determine which principles are applicable in a particular context and ensure that these guidelines translate into action in the development of the AI-enabled system (Halme et al.,

2021). As explained earlier, most research on AI-enabled PF design focuses on improving the functionality as a prediction tool or metric to measure passenger flow volume. A study by Leikas et al. 2019 outlines the need for a value-centered approach in the ethical design of autonomous systems; however, it is not specific to AI ethics (Vakkuri et al., 2021). Therefore, there is currently no research on using ethical development tools like ECCOLA to generate user stories that can aid the ethically aligned design of an AI-enabled PF system.

## From User stories to Ethical User stories

In software engineering practice, user stories are an industry driven research (Agile Development) and shares some academic research of the area. Practically, user stories are a few sentence lines of high level customer requirements formed by the customer or by the developer team. They are written on index cards, post-it notes or predefined template that follows the same structure, which is elaborated below in more detail. Sometimes a product owner is named by the developer team to form user stories. These software requirements define the functions that are needed for the software to work. Non-functional requirements define the needs that are vital for the software to work e.g. reliable, and that it is experienced usable, consistent and safe to use, and among others, it is ethically aligned. Developer team creates many user stories, where ready user stories are prioritized and the most valuable user stories towards the customer, are used in the first sprint of software development, where they are picked from the product bag log. (Cohn, 2004) When the non-functional requirements define the need that are vital for the software to work e.g. reliable and safe, themes shared with the ECCOLA method, we can exploit the non-functional requirements categorization towards Ethical User story practice and describe that ethical user stories relates to non-functional requirements.

In detail, the user stories are divided into two main sections. First section describes the particular requirement which is formulated: "as a [user], I want to [capability], so that [receive benefit]". (Cohn, 2004) Second part will offer the resolution for the requirement: "What needs to be done". (Cohn, 2004) As an example of Smart terminal Ethical User stories:

> *"As a data analyst (or the appropriate role) using the bought data for the system, I would like to know that the data are of good quality and not biased so that I can provide an effective and unbiased service for users of the system."*

"What needs to be done", part is a customer driven (acceptance) tests that gives the customer requirement a practical mean to form a real-world software feature. Sometimes the agile requirements team uses a ready formula for the acceptance phase that follows pattern of, [Given][When][Then], which gives the acceptance test a context, action to be carried out and consequences to be obtained. (Agile Alliance, 2021) For the given example above, the acceptance criteria could be:

> *"Documentation and logs of data sources for traceability are kept."*

## 3. Research Framework

The research was conducted in six (6) month timeline in 2021. Several activities and group sessions were arranged in that time period to collect the empirical material. Raw data, consisting of 300+ notes, for producing ethical user stories were generated through three (3) workshops during the summer 2021, held by JYU AI ethics lab. Participants invited to workshop were mainly from industrial and institutional partners and assigned to SMARTER Work Package (WP) 1 - Business Transformation, production.

In general, the project SMARTER is divided into 5 work packages, progressing sequentially, with all having their own targets. Ethically aligned design concentrates in the initial project phases to give the developers and participating stakeholders the opportunity to follow and cover ethical requirements in their design towards the following project phases and for the replicable model blueprints. The leading company had divided the development work in all work packages into three (3) use cases: Ship Turnaround, Truck Traffic and Passenger Flow (PF). PF is further polarized into two sub use cases: PF with and without a vehicle. As mentioned earlier this article is the first publication from the series of three and in this article we concentrate on the passenger flow with and without a vehicle.

In practical terms the use case passenger flow in the SMARTER project is targeting to the optimization of passenger flow in their end-to-end journey by improving different sub-systems. According to the project documentation these are e.g. digital customer experience like virtual check-in, public safety as in utilizing situational awareness, and video surveillance. MAAS -concept (ERTICO – ITS Europe, 2021) is also one area of added value in optimized passenger flow through mobility service, where person transport (e.g. cars, collective transport, bikes) and events are connecting the ferries.

The process of making user stories based on Ethically aligned Design, using the ECCOLA method was multi-staged, evolving ad hoc. The three (3) use cases were managed in three (3) virtual workshops and were individually processed. Use case owners were asked to choose themes to suit best in their context use case. In Passenger Flow the themes were: Transparency, Data, Safety & Security and Accountability. The chosen ECCOLA cards were: # 5 Traceability, # 7 Privacy and Data, #12 System Security, # 13 System Safety and # 18 Auditability. Cards were placed on an on-line whiteboard for reference. Each workshop took 1,5h to 2h, where the cards themes and topics were processed separately in 5 sprints, every sprint taking around 10 minutes. To enhance discussion participants were divided into break-out rooms with a facilitator from the AI Ethics lab. Participants did get an induction on the subject (Ethically aligned Design and ECCOLA method) and also instructions on how to make notes on the whiteboard. The idea of the workshop was to bring into attention any ethics related concerns or issues that needed attention or action in the project that wasn't discussed or didn't have a forum for it.

Participants made notes based on the discussion or based on their previous experience in the project and the notes were transferred to a spreadsheet to thematical classification. As the notes were previewed by one researcher, the overall experience of the review concluded into three (3) main categories for all the use cases: Human, Technology or Practicality and Data & information centric categories. This categorization helped the researchers to create use stories. AI Ethics lab personnel (12) were invited to the user story process activity and seven (7) researchers participated to the virtual activity to create ethical user stories from the workshop notes.

The Ethical user story development team consisted of 8 people, where 7 of them were PHD students and 5 of them working in the AI ethics lab as PhD researchers and the rest two students were acting also as a practitioner in the software engineering field and HCI design. One participant was doing his post-doctoral research. Six studied the field of information system sciences, one cognition sciences and one computer science. Two of the PHD students were first year students, two second year and three fourth year students. The Ethical User Story development team was interviewed at the end of the process and asked questions about the assignment and process /Appendix 2. The development team proceeded in several sprints and met several times online to proceed with the flowing assignment, meaning that the recurring meeting was taken as many times as the applicable notes were left on the spreadsheet.

Below is a table of the developers experience of in the field of software engineering.

*Table 1 Experience in the field*

|  | How much work experience in the field of software engineering / development do you have? | How much experience in Agile development do you have? | How much experience in writing user stories do you have? |
| --- | --- | --- | --- |
| None | 4 | 2 | 1 |
| Junior Developer (1-5 years) | 2 | 6 | 7 |
| Senior Developer (5+ years) | 2 | 0 | 0 |

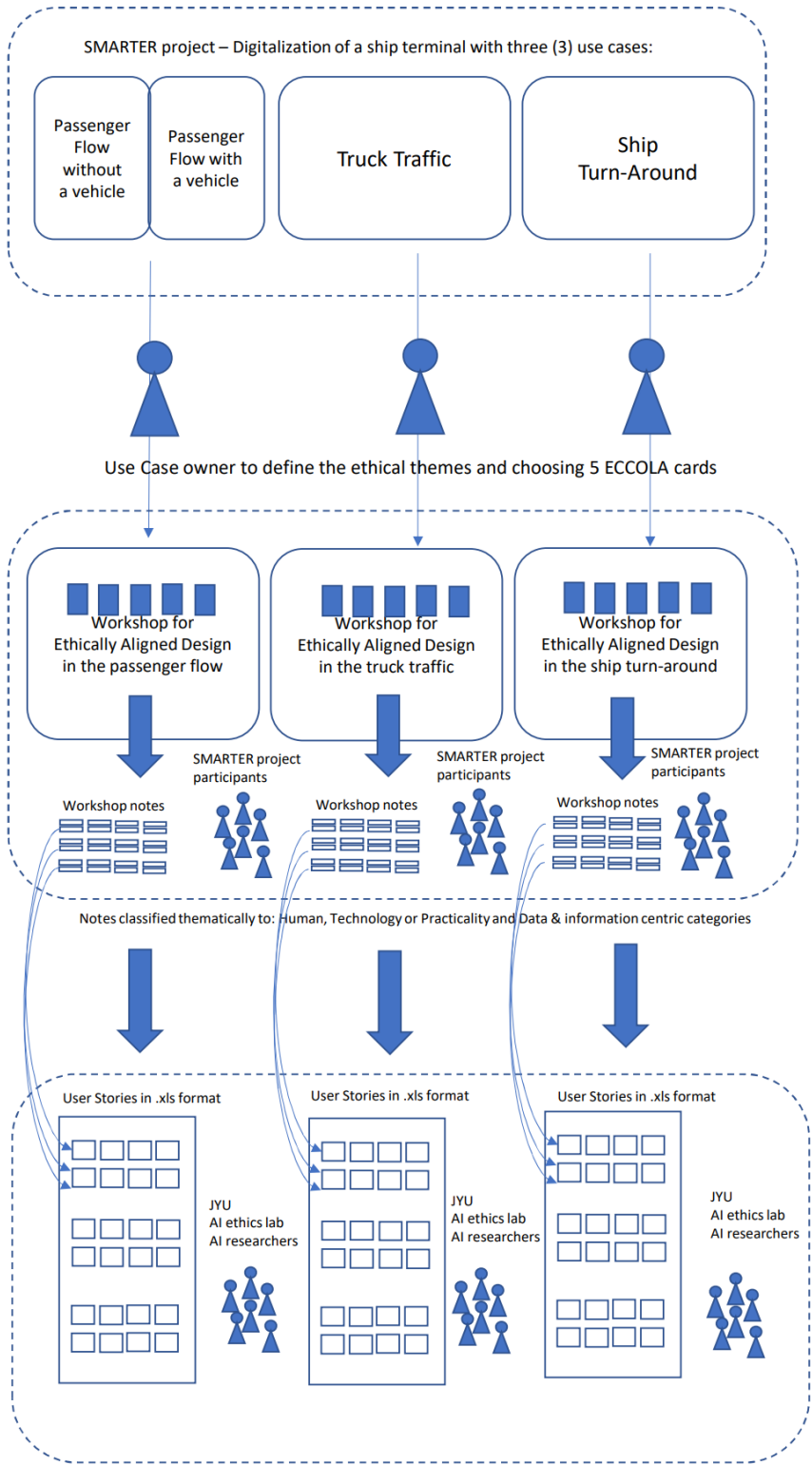

*Figure 1 Ethical User Story process in the context of SMART terminal -project*

## 4. Reporting the results

The format that the researchers used was according to the theory and is witnessed in the walkthrough of Ethical User stories /Appendix 1. In the first column is the raw data followed from the workshop authentic notes that prompted the user story development, in the middle the actual user stories following the format in practice: As a [user], I want to [capability], so that [receive benefit]. On the right column is the Acceptance criteria, also following the format in practice: [What needs to be done]. Example of each card theme is reviewed in table 2.

Not all of the notes from the workshops could be used due to their vagueness in context or unclear content. Altogether 65% of the workshop notes were applicable to create ethical user stories from all the passenger flow use case. 125 User stories were made out of 193 notes.

*Table 2 Example of Ethical User Story in ECCOLA theme Transparency and card # 5 Traceability*

| It should be transparent from the customers' and port's point-of-view - why their car is being processed by the port and what information? | As a passenger, I want to know exactly what information about my vehicle and implicitly, me, is being processed by the port, so I can feel safe and informed. | Before a person with a vehicle enters the port, they are given explanation of what information will be handled, for example at the stage of making a reservation when they agree to give the necessary information of their vehicle |
| --- | --- | --- |
| Why are you asking my data and for what | As a passenger using the terminal services, I want to know what data is collected of me and why, so that I can feel secure that my information is safe | A detailed document explaining the extent of the system's data use should be prepared and available to customers upon request. |

Overall feeling about the development of user stories was that it was very effective way of doing, the conversations taken after each sprint was good, but it was challenging to write user stories when the context was unknown, but on the other hand it gave freedom to be more creative when the mind wasn't set into any particular environment or scenario.

Participants also felt that it was interesting to see the further steps in the process, even though they felt that sometimes it was not that ease to produce user stories out of the workshop raw data due to notes' open-ended character. There might have been only one word, which didn't give any spark to develop from.

Participants did feel that it was a great avenue to learn and the process was ready set up, even though little bit ad-hoc in character. Interesting was that some experienced that when the process was ready set up, it took the pressure out of making ethical user stories and was considering doing "plain" user stories, which helped the development work. Also it was much more understandable see it in practice as one of the developers expressed: *"Being a participant of developing ethical user stories, it was much easy to understand that how exactly someone could put it into practice."*

Some felt that the process was unfinished as the validation of the user stories were not mentioned during the activity. The participant felt that the validation on the user stories should be left to the experts on this field and in this context to the project industrial or institutional partners.

This research was searching to answer to the following question of "How to write ethical user stories?" and What are the ethical requirements in Maritime industry, especially in port terminals when switching over to SMART terminals? The Ethical User Story practice, meaning the process, where the ethically aligned design tool ECCOLA was used in the research framework, empowered and released the developers from the heavy burden of ethical consideration and aid the developers to make user stories from project requirements, which was ethically aligned design and requirements for the project. Through the Ethical User story process the project received research based recommendations on business, data usage & sharing and many more with ethical consideration. This is again witnessed in appendix 1. How was it made? Based on the study results, we can discuss that the Ethical User Story practice is proof-of-concept for ethically aligned design practical tools.

## 5. Conclusions and Discussion

Practitioners and industry are at this point designing, developing, deploying and using SMART systems globally. What are the implications of those and are they ethically aligned or assessed? Now there exist a proof-of-concept practical tool, Ethical User Story tool, that are experienced as easy to use, and relieves the developer from the intractable and vague concept of ethics to enjoyable journey of learning and efficient development. This serves more attention and of course more cases for research and we are now calling for more participants to involve into this development work to get more results and research interest on it. This will continue with the following similar publications of other two use cases, truck traffic and ship turnaround, but also other prospects and publications in the future on Ethical User Stories.

# Appendix 1

# Passenger Flow with a vehicle

# Transparency

# 5 Traceability

Human centric (e.g. training, requirements, needs, goals, regulations)

| Traceability for boarded people in the vehicle+ vehicle | "As a security officer who got contacted by the police about a certain car, I want to track where the car was seen last, so that I can predict where it is going next. | An intelligent system connected to the video surveillance in the terminal can be given parameters to search for the current or latest location of a vehicle or person in the terminal by being given the register plate, a photo of a person, or other such identifying data. |
|---|---|---|
| | As a security officer who got contacted by the police about a certain passenger, I want to be able to get help from an AI system in tracking down a certain face, so that the person can be stopped within the terminal." | |

| If anything goes wrong, that I can find help at any stage to rectify issues | As a passenger with a vehicle, I want to have a helpline information (phone number etc.) that I can use when something goes wrong. | Terminal area and booking documents and booking app contains the help information (for different purposes broken down). |
|---|---|---|

| It should be transparent from the customers' and port's point-of-view - why their car is being processed by the port and what information? | As a passenger, I want to know exactly what information about my vehicle and implicitly, me, is being processed by the port, so I can feel safe and informed. | Before a person with a vehicle enters the port, they are given explanation of what information will be handled, for example at the stage of making a reservation when they agree to give the necessary information of their vehicle |
|---|---|---|

| Managing the complexity of traceability and GDPR | As a technically oriented passenger, I want to know and have the data available only for me of what data of my car is handled by AI so that I know the decisions concerning my car is made by AI and not a terminal personnel. | A clear advertisement is placed on the booking document about the AI and data privacy. |

| Crisis management for instances in which everyone wants their information erased | System administrator I want to manage the GDPR requirements in efficient manner so that the service is not jammed by the user information request or erase request | Data information policy is communicated at the service app/web page. Procedures in place for GDPR compliance and required steps are planned for example huge amount of user data request. For example redirection of resources or waiting times are clearly communicated. |

Technology or Practicality centric (e.g. testing, back-up, systems, logistics)

| External systems related to vehicles might be need to take into account | As a driver of a modern vehicle, I want to make sure that no port system will confuse or abuse any other system in my vehicle, e.g. bluetooth or self-driving systems, so that I can still trust my vehicle when entering the port | No feature of the smart terminal or port must interfere in an unwanted way with any system in the passenger vehicles. |

| If there is automation how is the car indentation done and how are errors handled | AS a driver, I want to know how to drive into the ship to right place so that the ship is stowed efficiently, and where all the booked vehicles can fit without any problems. | Car indentation system (digital, semi-automated, etc.) with control, navigation and markings is placed in the car deck to direct cars in line and to right places. |

| How to count people in the car | As a port official, I want to be able to reliably and automatedly know how many passengers are traveling in one vehicle, so that the border control is secure | The system has a reliable way to detect people sitting in a vehicle, that is in balance with respecting the privacy of passengers and adhering to border protocols |
|---|---|---|

| traceability for the flow from un even passenger count | As a port official, in the case of a seemingly failed passenger count performed by the system, I want to be able to track the possible time of the error so that the cause can be found and the error corrected | The system is transparent in its functioning and marks possible "uncertain" situations with timestamps or other identifying information |
|---|---|---|

| traceability for the flow from un even passenger count | As a frequent sea traffic passanger with my own car, I want to know that my car is not mixed with some other misbehaving (e.g. criminal etc.) drivers car, so that I can trust the system and continue to choose the company that I have used to. | Car identification system is tested with right and wrong data and error handling is defined in the terminal action plan and terminal personnel is trained with error handling |
|---|---|---|

Data & information centric

| Can the passengers trace their data? Do they have access? --> right to access and right to erase according to GDPR | As a passenger wanting to know what data about me is handled, I want to be able to get access to my data held in the system and get it erased, so that I have autonomy and agency over my data. | The data tied to a person can be found and is eraseable in the system upon request. |
|---|---|---|

# Data

# 7 Privacy and Data

Human centric (e.g. training, requirements, needs, goals, regulations)

| What is within the parameters of the terminal's responsibility | As an official of safety, I want to know, who to contact in terms of misbehaviour so that safety is ensured in all cases. | Smart terminal area is defined in the blueprints of the SMARTER project with responsibilities and roles. |
|---|---|---|

| Could there be privacy problems | As a passenger, I want to make sure that my data is not unnecessarily stored in the system so that it can't be misused or stolen. | The data storing and collecting follows the GDPR regulations and identifying data is anonymized or destroyed within the legal time frame. |
|---|---|---|

| What happens if passengers deny permission? --> deny access to service? | As a booking site manager, I want to make sure that the passengers have agreed to the terms and conditions before booking their journey, so no laws are violated when they enter the terminal. | Before being able to finish purchasing tickets, the passenger is required to accept the terms and conditions of using the smart terminal and having the minimum required data collected. |
|---|---|---|

Technology or Practicality centric (e.g. testing, back-up, systems, logistics)

| Providing arrival time information for the cars | As a car stevedorer, I want to give time information on digital boards or on apps how the ship stowage is proceeding to avoid multible inquiries on the subject | Communication plan has a section about on-time communication for different logistical purposes towards car carriers |
|---|---|---|

| there might be applications which utilize vehicle information and are not related to passenger flow | As a vehicle owner, I want to know, where my vehicle information is used so that it is not used in matters where it doesn't have my consent. | Data consent lifecycle is defined in the SMART terminal blueprints. |
|---|---|---|

Data & information centric

| What new data does the car provide? | As a former passenger without a driver's license and now with a lisence I want to know and see some visible information about the flow of events with the car when entering to the harbor because I haven't experienced the passenger flow with a vehicle before and I'm a little bit nervous. | Harbor area will have clear signs of the data requirities and visualize the flow of events. |
|---|---|---|

# Safety & Security

#12  System Security System Safety

Human centric(e.g. training, requirements, needs, goals, regulations)

| Humans are not good at following instructions | As a passenger, I want the process to be as simple as possible, so that I can smoothly board the ship without having to worry about following complicated instructions in a potentially high-pressure situation. | The boarding process is as similar to what it used to be as possible, or even simpler. Any and all instructions are crystal clear so that the passengers are able to proceed without asking for assistance as often as possible. |
|---|---|---|

| Delay can cause people to the actions to their own hands | As a border control personnel, I want to be prepared for delayes with the ship turnaround so that I know what to do with passengers behaviour in those situations. | Action plan for personnels in terms of delays are embedded into the logistics plan and it will have a section on human behaviour with smart system delays and actions preventing misbehaviour on passsengers. |
|---|---|---|

Technology or Practicality centric (e.g. testing,back-up, systems, logistics)

| Overbooking? | As a passenger with my family in our vehicle, I would like to know the capacity of the system before hand so I can avoid a situation where it is overbooked so that I do not turn up and not be able to use the service | The system can display capacity information for passengers that access the service remotely and at the port, display terminals can also provide these types of information for passengers. |
|---|---|---|

| System integrations to harbor services | As a system administrator, I would like the system to be easily integrated so that passengers driving into the port can easily access information | Clear logs that shows the state of the system and provide appropriate feedback and APIs that can improve system integration. |
|---|---|---|

| Can normal cameras capture everything? | As a surveillance system developer, I want to know the parameters and attributes of the cameras so I can develop a surveillance system that security can use and rely on. | Technical data of the camera is communicated to developers and also to wider audience in order to built trust on the system as a whole |
|---|---|---|

| vehicles add more external systems which are connected to the 'terminal'/passenger flow system and introduce more work to cover system security aspects | As a surveillance system developer, I want to have the breakdown of different use case ascenarious of the system so that I can develop the surveillance system to serve peoples safety. | Different use cases in passenger flows are modeled and communicated to all needed participants torugh the development lifecycle. |
|---|---|---|
| If there is an accident, what happens? Cannot read plate numbers? Snowy? Someone blocks platenumbers... | "As a passenger, I want the new system to not make boarding any more inconvenience for me, so that I can board the ship with my vehicle just like before (or hopefully even faster). | Passengers are instructed to clear their register plates before arriving to the terminal. An employee is still present to resolve problem situations (e.g. to quickly clean the register plate if needed). Accidents are handled as they currently are when two human drivers are involved. For accident situations with smart vehicles, accountability is also determined by the smart vehicle's accountability. |
| | As a passenger, I want it to be clear what should happen if there is a problem situation, so that I can board the ship with my vehicle just like before (or hopefully even faster)." | |
| Passengers with smarter vehicles: potential risks | As a passenger with a smart vehicle, I want to make sure that the systems in the terminal don't maliciously affect my vehicle and vice versa, so that I can use the smart terminal safely. | If the terminal interacts with smart vehicles, it must be in a way that does not compromise the safety of either party. There must be clear limits to what data is shared and what permissions the systems have for information exchange. |
| How does the smart harbor handle smart vehicles | As a passenger in a Smart vehicle in Smart harbor, I want ot be sure that I manage to get into the ship safely with my vehicle so that I can go and book a trip for my whole family. | Smart harborfollows a protocol on traditional cars and a smart vehicles when developing the flow of events from harbor entering to entering to the ship. |

# 13 System Safety Audi

Technology or Practicality centric (e.g. testing, back-up, systems, logistics)

| how to handle the lithium batteries | As an EV driver, I want to know, If I need to take any extra procedures on my car before entering to the terminal area or to the ship. | Boarding pass on cars are shared with instructions for different vehicles. |
|---|---|---|

| Failure (traffic light etc.) - failure to recognise the right vehicle --> what happens to ship loading or unloading? | "As a passenger driving his car at the habor with my family, I would like an alternative system in place if traffic lights fail so I am not stuck in a traffic jam with my family as this can cause stress. | The system can provide a standby system in place should the traffic lights fail or a human traffic warden that will help direct traffic for ship loading and unloading. |
|---|---|---|
| | As the traffic operator in charge of operating traffic lights for ship loading and unloading, I would like an alternative system if traffic lights fail so that I can continue with ship loading and unloading." | |

| When you get off ship there are police and check-points, for alcohol --> ways of assisting police? | "As a passenger coming off the ship I would like to be quickly checked by the police so that I can disemabark the ship on time | Digital scanners that synchronize passenger information with regulations such as alcohol limits that can be brought in and to ensure that drunk or underage passengers are not driving. |
|---|---|---|
| | As a police officer checking passengers at the check point, I would like a system that helps me quickly and efficiently to scan passengers ." | |

Data & information centric

| From the safety perspective - we want to know that we are getting the right, accurate information | "As a passenger with a vehicle, I would like to know that I have accurate information for me and my vehicle so I am not misinformed or violate any regulations. | Regular digital or even manual information system checks with prompts informing on current updates such as parking information, duration of parking, etc. |
|---|---|---|
| | As the information officer, I would like the system that provides information to passengers to be always up to date and accurate so that I can provide accurate information to the passengers and thier vehicles to avoid any stress loss of information can cause." | |

| Means of verification - authentication of information | "As a passenger with a vehicle, I would like me and my vehicle to be authenticated so that I can leave harbor when I have to and not have any cause for a case of mistaken identity. | Digital verification devices such as number plate scanners for vehicles to ensure that vehicle is authentic and in line with regulations and biometic scanner for human passengers. |
|---|---|---|

# Accountability

# 18 Auditability

Human centric(e.g. training, requirements, needs, goals, regulations)

| Vehicles increase work on auditability, due to vehicles and systems needed for those locally and externally | As the person in charge of registering all the vehicles that use the port, I need to be able to accurately record how many vehicles register online or in person and to reconcile that with the number that make the trip. | APIs and Interoperable systems that can provide accurate data records from different teriminals and can synchronize efficiently with the ports system |
|---|---|---|

| Is the booking different? | As a passenger driving into the port, I need to know that the information that I get from remote terminals is clear and thesame as the one provided at the port gate so that I avoid any ambiguity or wrong information | System provides clear and transparent information without any ambiguity at remote terminals and at digital terminals and with people who check in passengers at the port. |
|---|---|---|

Technology or Practicality centric (e.g. testing,back-up, systems, logistics)

| Attention to be placed on data management system - can AI be used for identifying relevant information for relevant parties? | "As a passenger with special needs and my vehicle adapted to suit me, I would like pertinent information for me and my vehicle to aid my use of the harbor. | AI integrated into the system that can help identify data patterns for users with special needs and provide necessary information. |
|---|---|---|

Data & information centric

| Can effective data consent systems be HCI of the future? --> can consent span the entire passenger experience? | As a driver of a SMART vehicle using the terminal, I would like a situation where the system can connect with my vehicle and communicate my data consent to me using voice and for me to give consent in thesame manner. | A clear and well defined HCI system where voice can be used to communicate data consent in line with relevant regulations. |
|---|---|---|
| Its not just the terminal operator - border control, guard, police, passengers --> data should be available for all these parties | As a user of the system in any of the capacities mentioned, I would like an aggregated system that gives me access to relevant data or information that I would need to carry out my duties | A centralized data management system with clearly outlined details on access rights and clear logs on users. |
| Passenger information: when booking for truck journey, driver's info is not confirmed, unlike in passenger traffic. | As a safety officer at the SMart terminal, I want to be aware that all passenger are identified with appropriate identification documents, so that I can do my job properly and responsible. | Booking information does not exclude anyone from the needed identification information and the booking application has the identification information as a mandatory field. |

# Passenger Flow without a vehicle

## Transparency

### # 5 Traceability

Human centric (e.g. training, requirements, needs, goals, regulations)

| Avoiding risks brought up by bias | As a passenger, I want to have smoothly pass at the security check, so that I will not be stopped because of my race, gender, age, etc qualities | Ensure the quality of the data by testing. Testing at least with race, gender, age should be done and variance between gropes should not be significant. Realtime testing in real context needs to be done before launch. |
|---|---|---|

| Ability to show where we have come from and where have been. | As a passenger I dont want to get lost in the terminal so that I wont be late from my cruise | I while shopping in the termina I lost track of time and I need to locate my self in the terminal. Mobile app or feature in terminal enduser app for navigation. System has Opt-in policy. Treminal has tracking points, sensors or QR code to provide tracing of the users movements and provide indor navigation to desired points. |
|---|---|---|

| User org. needs to have doc | As a system operator I want to documentation of the different components of the smart terminal system so that I can explain system properties in auditing | When auditing is in process system need to provide detailed documentation in every level of system architecture. Also 3rd party components needs to be documented or at least links to those systems documentation needs to available |
|---|---|---|
| Do I understand where I am in relation to the rest of the system? | As smart terminal service provider I want to understand my services relation to others so that I can integrate my service to the smart terminal with out causing chaos or service don't time | System needs to provide process documentation and specs for new service joining the system. List of acceptable actions should be given. E.g. are there restrictions of data usage, bandwidth, number of API queries… |
| Relevance for the systems user | As I annoyed User I want to understand why the terminal apps are showing me information of the service traceability so that I can be not annoyed or get rig of the popups | System needs to provide user friendly and short description of the traceability features of the service and also provide instructions to take acting to access traceability. System needs also be able to end user notifications on traceability if user wishes so |
| User org. can define demands | As Smart terminal service provider I want to have traceability on the smart terminal services usage and user demands so that I can focus service development efforts | Smart terminal system need to have API for BI tool so that the service provider can track the service usage. |
| Demand for traceability | As a government official, I want to trace the actions of the smart terminal, so that I can find out if its actions are acceptable | System should create explicit action log of its actions. The logs should be accessible for designated personnel and understandable or reliably interpretable for government or other officials who require the logs for a legitimate reason |
| user exclusivity (foreign and elder people e.g.) | As terminal elder aid provider I want to track the aid resources and comming customers so that I can direct required help where need | Systems identifies users with special need and has feature to inform terminal service provider of required help. Systems also need to be available to tract special need aids inside the terminal |

Technology or Practicality centric (e.g. testing, back-up, systems, logistics)

| | | |
|---|---|---|
| I.) AI functionality should be based on requirements and specs, so it should be transparent information on how AI works and why it does what it does. | As an employee trying to fix an error of the AI system, I want to be able to find out why the AI did what it did, so that I can correct its behavior | AI system needs to have a function to generate an explanation for its actions if requested by a system administrator with rights to access the system |
| II.) According to normal SW development process. | As an employee trying to fix an error of the AI system, I want to be able to find out why the AI did what it did, so that I can correct its behavior | Function level documentation needs to be in place. |
| III.) You should have versatile use cases, which you are validating at the end. | As software developer I want to test my system with multiple use case so that I can reassure the client that the system is doing with it says it does. | System needs to be tested with given number of versatile use cases and not only obvious use cases with acceptable results. Also, those cases that would be possible but not testes should be clary stated |
| Traceability for price | As a customer, I would like information on freight prices to enable me plan my budget | When planning my budget for freighting my goods, i would like clear information on the different pricing options (aggregated prices) maybe in an app, sites or displayboards at the terminal to help me plan my budget effectively. |
| Pricing criteria | As a passenger who uses the booking system, I would like clear and well defined pricing criteria to give me the necessary information i need. | clear and well defined pricing criteria such as pricing analysis for customers on the system |

| Interaction between physical and virtual spaces - how can the user orientate and be assisted through | As a customer, (older customer), I would like to know if I am relating to a human or machine to help me know how to channel my inquiries | When making inquiries as a customer, are there pointers to say you are communicating with a virtual assistant and is there provision for a human assistant where possible? |
|---|---|---|

Data & information centric

| Data traceability | As auditor I want to understand what kind of data was used for the system so that I can be sure the data is representative for the given use context | When asked system documentation need to have description on what kind of data and from where was used to teach the system. Data set with out clear description should not be used |
|---|---|---|

| XAI | As system developer I want to understand how the AI component works so that I can understand what causes the error in the system. | In service critical or human relating features only explainable AI/ML solutions are used. When requested the system can provide reasoning behind given result. |
|---|---|---|

| Needs to be transparent for the users - what the software needs to do; information from the people, process and give outputs | As a passenger using the terminal I want to know that the data collectd from me is used for the purpose i provide the data for | When prompted by passengers, the system should be able to provide information requested, eg customer asking for summary on a trip should be able to get one from the system |
|---|---|---|

| Ethical aspects should be taken into account with GDPR, prior any implementation and then separately in each use case. | As an administrator of the system, I want to make sure that the system aligns with GDPR and the data is handled responsibly, so that the system is functioning legally | Before implementation, there should be a way to verify that the system is following the best practices and aligned with required legal standards |
|---|---|---|

| Why are you asking my data and for what | As a passenger using the terminal services, I want to know what data is collected of me and why, so that I can feel secure that my information is safe | A detailed document explaining the extent of the system's data use should be prepared and available to customers upon request. |
|---|---|---|

# Data

# 7 Privacy and Data

Human centric(e.g. training, requirements, needs, goals, regulations)

| How to provide data collection policy info | As a passenger I want to have clear and easily understandable policy notice on what personal data is collected from me and for what purpose and that I can opt in/out of "non-vital" policy info, if using an app with extended services like tracking my movements in the terminal. | The user is offered an easy to read statement of data privacy policies excercised with the terminal services for the services she is using (off the app) . She can opt in and out of data collection not necessary for my need of services. |
|---|---|---|
| Dilemma between regulation and what is being collected | As a privacy officer in a stakeholder company joining the smart terminal, I want to make sure that the system is only collecting data it is allowed to, so that my company does not get any legal repercussions | System provides Data collection description and Data restrictions. Relevant contact information. Way to notify errors |
| Can data be sell to others? | As a passenger, I want to know if the data stored of me in my use of the terminal can be sold to third parties, so I feel secure with my agency over my identifying information | Clear security communication to passengers who intend to use the terminal services, about what data is collected and how it is used |
| Anonymization and other measures (e.g. providing video surveillance in restricted area and during restricted time) are recommended to overcome GDPR violation possibilities. | As a passenger, I want to make sure that my personal identifying data is not available to everyone in the system so that I can feel safe it's not being misused | Video recordings and such unanonymized identifying data are only available to few certified people |
| Information on who has access to user data, is it used for marketing, statistics? | As a data privacy officer (or similar role) I want a registry of use purposes for different personal data "contents" is maintained and use of data is explained by the requirements set for data privacy notice and this information is maintained so I can ensure that the personal data use cases are communicated for the end user and their concents received, where applicable. | A register of personal data processing activities and their justification (basis for processing) is created, all involved parties are identified and this register is maintained and made available for needing parties. |
| informing users and giving them options | As a passenger I want to have one "info sheet" of data privacy, which is easy to read and not too long, so I can understand what purposes the data I provide is used for. | The data privacy notice is made available to the passenger (via the app she is using), which provides the purposes the personal data is used for. |
| GDPR clearance needs to be done in the beginning of the project | As I use case coordinator I want to understand of the systems and components contributing to the smart terminal system so that I can asks if GDPR clearances are needed and have the relevant system provide dealt with them | Use case coordinator has detailed plan of the systems and components contributing to the smart terminal system and also disclosure from those parties if they need GDPR clearances. |

| Control over data | As a passenger I want to have control over the data which is not fundamentally "minimum needed for the travel service", so I can opt in or out of data collection per my preference to share data with the operator. Later I want the option to "delete" the data, which the operator has no legal basis to maintain, as per GDRP data subject rights. | The value add application the passenger is using needs to impelent a control which identifies, which data processing activities are necessary for which basic or added value services the user is considering to use and she can thus opt in or out of the service based on the data processed. The data the user has selected for her rights execution needs to be processed accordingly, e.g. anonymized, deleted or made available for the user to transfer to another service/download. |
|---|---|---|

| Why is the data collected? Is it for safety and security of the passengers. | As a VIP passenger I want to understand why is the data collected so that I can trust my VIP data with the system | Data statements need to be available and also descriptions what services optional and how to opt—out on those need to be implanted. Description of what parts of the system are used by harbor official and cannot by opt-outed |
|---|---|---|

| IF there is a risk for data breach, that needs to be clarified in DPIA and end users need to be informed on possibility for GDPR violation. | As Concerned smart terminal data security office I want to cover possible risks in DPIA so that I can be sure that privacy and data breach has been considered by the system providers | DPIA must be made to meet the given standards. Relevant stakeholders for the SMART terminal system need to be informed about the required level of standards for Data and privacy when operating as a part of SMART terminal system |
|---|---|---|

Technology or Practicality centric (e.g. testing,back-up, systems, logistics)

| Data layers? | As enthusiastic sw developer I want understand the SMART terminal systems data layers so that I know on which layer my service can operate and what data sources are available or me | System architecture needs to be draw and it needs to be shareable with relevant stakeholders. Possible security issues relating to architecture information sharing needs to be mitigated |
|---|---|---|
| Passenger flow video, security, immigration, transit hall shops, transit hall wifi? | As a passenger using the terminal I want to be assured that all the data gathered on me from all these sources are used in line with GDPR and not used out of context | The system can notify users on their tickets and display boards that data being collected are used in line with GDPR. Also regular tesing and security checks to ensure that customer data are not being jeopardized |
| Mask detection - everyone wants another process: what biometrics can be used around a mask? | As a passenger using the terminal and I have religious beliefs where I have to cover my face, and now with the use of masks I want to know that the system has another way of identifying me besides me showing my face | Other forms of biometrics such as iris or finger print recognition as an alternative to facial recogniton |
| Network penetration tests aretypically required to ensure sufficient cyber security of the system, to minimize data breach risk. | As a cyber security officer, I want to do everything to ensure that the system is secure, so that our data does not get stolen | A regular testing protocol should be in place, to ensure the security of the system and protect against threats. The system should be kept updated and its resilience tested in regular intervals |
| How to accurately capturing data? | As an administrator of the system, I want to ensure the data processed by the system is accurate, so that the data is good quality for several purposes | Protocols for verifying that the system collects data accurately should be in place. The data accuracy and quality should be evaluated on a regular basis |

Data & information centric

| | | |
|---|---|---|
| What kind of data does the smart harbor produce? | As an administrator, I would like to enure that the data the habor produce are accurate to provide safe and economical services for cargo and passenger travels. | Ensure that channels used for data collection document processes to aid accuracy |
| What data would be valuable from a company and marketing point of view? If it improves the service offered, it may be fine. | As an administrator, I would like transport system-related data to help improve the transport services of the smart terminal | Secure APIs to help enable the process and open data sources |

| What kind bought data would benefit smart harbors | As a data analyst (or the appropriate role) using the bought data for the system, I would like to know that the data are of good quality and not biased so that I can provide an effective and unbiased service for users of the system | Documentation and logs of data sources for traceability |
|---|---|---|

| What data is relevant for smart terminals? | As a data analyst (appropriate role), I would like to have data that helps with transport services and customer analysis to help me provide data relevant for the terminal | relevant Data sources (APIs) are open for data analyst from the system and not just generic data |
|---|---|---|

| there are also other regulations for viudeo surveillance, in addition to GDPR, which are related to personal data. GDPR clearance doesn't solve all issues with personal data, when video is handled | As a passenger using the terminal, I would like to know that there are regulations that govern how surveilance videos are used that will guarantee the privacy of videos I am in and only relevant users are granted access. | The system follows regulations on data privacy and ensures that data access is granted only to relevant personnel. |
|---|---|---|

| The way the purpose of collecting data also matters. If it is a long legal disc | As a passenger, I would like to know why my data is being collected so that it does not wind up being sold or appear somewhere I did not give my permission. | Passengers can be notified by the system at collection points such as on remote terminals, or if they use apps at point of feeding their data why their data is being collected |
|---|---|---|

| Data lifecycle | "As a passenger, I would like to know how long my data is kept by the system so that it is not kept indefinately as this will give me peace of mind. | Regular data checks embedded in system to notify appropriate personnel on lifecycle of data to ensure conformity to regulations |
|---|---|---|
| | As a data administrator(appropriate role), I would like to deal with data that are up to date or in line with regulations that have not exceeded their duration period as this will help me work with current and relevant data and at the same time let me know that I am not violating any regulations" | |

# Safety & Security

#12  System Security

Human centric(e.g. training, requirements, needs, goals, regulations)

| Guidelines for strengthening the system, system needs to be monitored for vulnerabilities; pen-testing | As a passenger, I want to be able to trust that the system is a secure one, so that I will not be inconvenienced (or worse) by a cyber attack of some sort. | System is continuously monitored for vulnerabilities. Penetration tests have been conducted, and new ones are conducted as needed. |
|---|---|---|

| Keeping the human in the loop. I. e ramps to ships | "As a passenger, I want to still have some humans (from the company) around, so that I can talk to a human in case of problem situations / feel like the company cares about me. | The extent of necessary human oversight has been determined. The employees in charge of overseeing the process have been appointed. |
|---|---|---|
| | As an executive, I want to have a human in the loop, so that the employees can resolve any | |
| | unforeseen issues and execute a contingency plan if needed." | |
| integration of several systems operated by several actors is challenging | As a CTO, I want the different systems that need to be integrated together to work seamlessly, so that the system functions correctly. | Systems have been integrated successfully. |
| Adversarial attacks on neural networks | As a CTO/CSO, I want the system to be safe not only from conventional cyber attacks but also cyber attacks specifically targeting the AI/neural network components of the system, so that system is secure against all types of attacks. | To protect against data pollution, e.g., disable the system from learning in use (offline). This is an on-going issue where solutions are evolving right now, and various measures should be considered. |
| Different security risk and requirements for different parts | As security auditor I want to have descriptions of different security risk and mitigation plans so that I can validate the system | Detailed security plan from architecture level to single system feature level is drawn and mitigation plan is in action |
| Planning and processes before security breach - PRE-INCIDENT readiness | As a security officer, I want the staff to know what to do in case of a security breach, so that they know how to minimize damage and react fast if an incident occurs | An action protocol should be prepared for the occasion of a successful security breach. The system should be tested and kept as invulnerable as possible, but the staff should also be trained to take action to minimize damage in case of a successful attack on the system |

Technology or Practicality centric (e.g. testing, back-up, systems, logistics)

| | | |
|---|---|---|
| People flow from a security perspective - AI for physical security of hardware etc. | As smart terminal service provider I want to use new cool AI solution safely in the terminal so that we can better manage people flow from a security perspective. | People flow management related systems and hardware specific security risk are recognized mitigated. Possibility of system failure scenarios tested and possible physical harm from these failures consider. Backup plan ready for full system failure |
| Providing chaos at the terminal | As terminal security guard I want to trust the system so It doesn't cause chaos in the terminal and make my work harder | Understandable description what AI system does and don't do described for people working along side the system. Insides of the people working along side of these systems heard and involved in the design |
| Border control | As a government official, I want to be sure that the digital system cannot be cheated to give illegitimate access to people through border control or that it does not deny legitimate passengers, so that human rights are not violated and the country will not face repercussions | All the digitalized aspects of border control need to be trained for as many exceptional or rare cases as possible, and a human officer should monitor the digitalized border control |
| If someone accesses the system - what happens next? How does it detect that someone has access to the system? | As the system's security officer, I want to be able to know who has accessed the system, so that I know if someone without the proper right has gained entry | A function that alerts the chief security officer when something unusual happens in the system and someone accesses it from an atypical source |

| It is often recommended to test/validate cyber security of the system by third party | As a passenger, I want to make sure that the system that saves my data is secure, so that my data will not end up in the wrong hands | Cyber security of the system should be varified by testing by multiple professional parties, not just one |
|---|---|---|

Data & information centric

| Network penetration tests are today's 'normal' when building any system including data, especially personal data | As a cyber security officer, I want to do everything to ensure that the system is secure, so that our data does not get stolen (repeated story) | A regular testing protocol should be in place, to ensure the security of the system and protect against threats. The system should be kept updated and its resilience tested in regular intervals |
|---|---|---|

| One aspect of GDPR is cyber security - making sure that collected data is not leaked, kept within frame of the network | As a privacy concious celebrity, I want to see from the terminal internet page or somewhere in the media that my travelling papers and information is kept safe in all matters so that I can travel where ever I want to not being afraid that somebody is stalking me | Privacy plan should be placed in the internet page and linked to social media for everyone to see clearly |
|---|---|---|

# 13 System Safety

Human centric(e.g. training, requirements, needs, goals, regulations)

| How effective is human oversight if there is a system glitch? | As a customer, I would like to know that i can have assistance or access to help when should there be any system downtime which may affect passenger flow. | System should have Help line number to call or help button in an app in the system so that responsible parties can either fix the problem or provide alternative assistance |
|---|---|---|

| Human oversight - Human operators and how this affects system safety? Authentication? Access? | "As a passenger, I want humans to be able to take over in case there is some issue in the system, so that I am not (too) inconvenienced (or worse). | Authentication measures for human oversight determined (e.g., physical access). Override/oversight measures tested for vulnerabilities and considered from cybersecurity perspective. |
|---|---|---|
| | As an executive, I want my employees to be able to take over in case of any issues that prevent the system from functioning normally, so that the terminal can continue operating even if we face issues. | |
| | As a CTO/CSO, I want human oversight processes to be considered from a security perspective as well, so that they do not become a safety/security issue" | |

| Physical safety of the people? | As a passenger I want to have an understanding on the general functionality of the system from the perspective of being able to get a concept of any safety concerns the system's use might expose me to. | The passenger must be presented with any physical safety concerns of the system, analyzed and provided with assurance that no risk exist or some estimaton on the risk. (E.g. system might be exploited to track some imporant person's movements and plan physical assault on her or track peaks in terminal usage to plan terrorist activities for maximum damage). Such a estimation/statement could be provided as a pop-up or part of info page /FAQ. |
|---|---|---|

| Trainings to be able to deal with the systems | As an employee in the terminal, I want to understand how the system works (and how to operate any parts of it that I may need to operate), so that I can confidently work together with the system. | Employees have been trained to a) understand the system as a whole, b) understand the new processes, c) understand in detail the parts of the system they are actively working with, and (if relevant), d) operate any parts of the system they may need to (if not normally, then perhaps in case of unforeseen issues). |
|---|---|---|
| Backup procedures for all automated processes | As a passenger, I want to be able to board the ship even if there is some issue with the system, so that no system issue can cancel my trip. | The number of employees that need to be present to take over in case of any unforeseen issues has been determined. The relevant employees have been instructed how to take over in case any part of the process automation fails. |
| Many system safety issues relate to system security | As a security officer operating the system, I want to make sure that passengers cannot accidentally compromise the system with an error so that the system can continue functioning | The system does not allow access to any vulnerable parts of the system to unauthorized personnel. |
| System should not lead passengers into situations where their safety is jeopardized | As a passenger, I want to know that the terminal system will not lead me into a dangerous situation so that I can trust the system and I don't have to rely on my own inadequate capabilities in navigating the system usage safely | The system is clear and intuitive to use, and leaves as little as possible for interpretation. It will not offer possibilities for passengers to do a wrong, dangerous or forbidden thing |
| The systems will fail | As an employee working in the terminal/on the ship, I want to understand how the system should be working in different types of situations related to my job, so that I can see when/if it is not working as intended. | Human oversight procedures determined (e.g. when/how an employee can take control of the system or its tasks, or even shut it down). |
| How much control do humans have in the system, and how able are they to detect a glitch in the system? | As an employee working in the terminal/on the ship, I want to understand how the system should be working in different types of situations related to my job, so that I can see when/if it is not working as intended | Intended system use clearly defined so that misuse can be determined. Expected outcomes/outputs determined for each process; employees should know how to expect the system to behave so they can determine if something is off. Human oversight procedures determined (e.g. when/how an employee can take control of the system or even shut it down). |
| Accountability - who's accountable? Matrix of accountability and responsibility --> is there agreement and consensus? Regulation and standards for accountability --> project-based | As a CEO, I want it to be clear who is financially responsible if the automated processes fail due to a system error, so that I know how to proceed in the aftermath of such a situation. | Intended system use clearly defined so that misuse can be determined. Liability established through contracts. |
| Accountability -> Monetary loss when people cant travel | "As a passenger, I want to be compensated monetarily if the company cancels my trip, so that I can trust them with my money. | "Procedures for monetary compensation (redress) are in place for situations where an entire trip is cancelled. |
| | As a CEO, I want it to be clear who is financially responsible if the automated processes fail due to a system error, so that I know how to proceed in the aftermath of such a situation." | Intended system use clearly defined so that misuse can be determined. Liability established through contracts." |

| Accountability defined beforehand | As an employee working with the terminal, I want to be able to find out who is responsible for another part of the system, so that I know who to contact if I need to clarify something | Documentation that clarifies the assigned responsible person for each part of the system |
|---|---|---|

| How much automation we can handle | As smart terminal service provider I want understand what level of automation is possible to implement into the smart terminal culture vice (socially acceptable) so that we don't drive away customers | User and feasibility studies done for the planned system. Testing and scenarios before building. Feasible places for system automation recognized |
|---|---|---|

Technology or Practicality centric (e.g. testing, back-up, systems, logistics)

| Functionality of the system must be validated from the safety perspective. (physical and 'virtual') | As family traveling with children I want to feel safe in the smart terminal physicaly and 'virtualy' so that I am comfortable taking my family to vacation | Smart terminal system is tested from passenger point of view. Versatile scenarios and users are tested. Viewpoint of children is also covered |
|---|---|---|

| Role-based access and rights - levels of detail to access. Camera views and playback | As smart terminal security officer I want know who has access to what data sources to that I can make sure that officials has access and only they have access to information relating to their work | Role-based access and rights implemented. Management and documentation of these roles done. |
|---|---|---|

| Backup options needs | As a passenger, I want to be able to board the ship even if there is some issue with the system, so that no system issue can cancel my trip. | The number of employees that need to be present to take over in case of any unforeseen issues has been determined. The relevant employees have been instructed how to take over in case any part of the process automation fails. |
|---|---|---|

| Continuous testing | As a cyber security officer, I want to do everything to ensure that the system is secure, so that our data does not get stolen (repeated story) | A regular testing protocol should be in place, to ensure the security of the system and protect against threats. The system should be kept updated and its resilience tested in regular intervals |

| Backup procedures for all automated processes | As a passenger, I want to be able to board the ship even if there is some issue with the system, so that no system issue can cancel my trip. | The number of employees that need to be present to take over in case of any unforeseen issues has been determined. The relevant employees have been instructed how to take over in case any part of the process automation fails. |

| AI should not never be 'left alone' to handle the system, unless the function is guaranteed to NOT to endanger someone's safety.. There should be always fallback loop, which alerts human to oversee/react the situation. | As a passenger, I want to make sure that my safety won't be endangered by an automation failure, so that I can feel safe during my travels | A human should surveil the actions of every AI system, particularly safety-critical processes/every AI system should have a designated employee who monitors its performance |

| Can you operate terminal manually | Question for the use case owner. Can you even operate current terminal manually? | |

| How can the companies handle SW down time | As a stakeholder of the system (booking service), I want to have clear protocols for how system down time is processed, so I can keep the trust of my customers | Backup options and system downtime protocols should be established before launch, and all relevant stakeholders should be informed about the downtime protocols |

| efficient and intelligent people flow management in 'panic' situations increases safety | As a passenger in an emergency situation, I want to receive good instructions so that I can get myself to safety without harming myself or anyone else | Staff should finish training for emergency situations, to calm the passengers and guide them in a pre-planned way to safety efficiently. The system should also be ensured to communicate safety instructions to passengers efficiently. |

Data & information centric

| efficient and intelligent people flow management in 'panic' situations increases safety | As a passenger, using the terminal, I can provide my data but I would not like the system to keep my data indefinately | Clear log and access report generated by the system on passengers from data collected. The system can also notify passengers on the period thier data will be kept while collecting the data |

## Accountability

# 18 Auditability

Human centric (e.g. training, requirements, needs, goals, regulations)

| Is the audition for the port system, service or components? | As an third party auditor I want to know how extensive is the audition plan so that I can do my job well responsible and offer safe environment for all stakeholders of the SMART terminal. | audition plan needs to be in place. System and its compnents are build and selected auditability in mind |

| audit requirements have to be in the system description from the beginning | As software developer I would like to know what auditability is in the case os this system so that I can meet the requirements | Audit requirements have been written out in system description. Requirements have been discussed among all relevant parties related to systems development |

| Auditability is also important from the GDPR point of view and hence it should be designed in the beginning of the project | As GDPR official I want to be able to see the DPIA plan so that I can determinate are further actions needed to assesses services for GDPR compliance | DPIA must be made to meet the given standards. If needed DPIA is aviabler Relevant stakeholders for the SMART terminal system need to be informed about the required level of standards for Data and privacy when operating as a part of SMART terminal system |

| This is linked to traceability, security and privacy - one aspect of GDPR> is that it needs to be auditable | Smart terminal provider I want to able to understand the systems as whole so that I can produce and ask for required documentation to meet the GDPR requirements | The system is designed and architecture is planned privacy and data protections auditability in mind. Clear documentation on how API and services joining the SMART terminal system will affect the privacy and data protection |

| Where are the lines of responsibility | As a service provider I want to understand what lines of responsibility, legal responsibilities and restrictions are so I can be confrontable on joining the smart terminal system. | Smart terminal system is a system of systems. The core system linking everything together needs to provide description responsibilities among systems linking to the system before service can come as partner to the system. Also requirements of information sharing and processing should be provided both ways |
|---|---|---|

| Well designed and modular system overall is good for auditing | As a CTO, I want the system to be developed using current best practices for software development (such as modularity) despite it being an AI system, so that it is more easily managed and more auditable. | The system is created from modular parts, to what extent possible. |
|---|---|---|

| Who is the right party to audit a system like this? | "As a CTO/CEO, I want to understand the audit procedures required by legislators, so that we can successfully deploy the system with the permission of the relevant government/regulatory stakeholders. | "The system is audited by 3rd party experts as required by regulations/laws.<br><br>The system is audited by 3rd party experts." |
|---|---|---|
| | As a CTO/CEO, I want to audit the system with a third party before we deploy it, so that I can ensure that it will work." | |

Data & information centric

| | | |
|---|---|---|
| Are there effective 'cleaning' or 'sorting' mechanisms for the data, so that relevant information is brought forward? | As a fast-moving sales person, I want to use a system that is easy to operate and see the information that I have chosen, so that I can move quickly from place to place and book cabins, food etc. easily | Developers should be aware that good usability instructions for developing are utilized when making the booking features / systems etc. |
| System data versus camera data - and what happens to the data? Data sorting is time intensive | As a an information manager of the SMart terminal, I want to know what kind of data architecture is needed and maintained thorught the data lifecycle so that we can buy/rent approbriate hardware or cloud place for our data. | Datalifecycle plan is made by the information management team for the developers to use from the beginning of the development phase. |
| AI that detects relevant data, for specific purposes | AS a quality responsible team member, I want to know on what criteria the AI chooses the relevant data so that I can explain to auditors the idea of it for them to make auditions in the future | Criteria list of different purposes detected from the information flow should be made and maintained |
| identity information is provided already in the booking process | As a customs officer or AI customs officer at the gate, I want to know that the one who has bought the ticket is really using it so that no violations against identity since the booking will be made. | The flow of identity checking should be simulated and tested against identity teft issues. |
| Log data collected from end-users - every click needs to be logged | As a CTO, I want the system to log any human actions, so that we can determine when/if someone made a human error (or did so on purpose). | System keeps track of all button presses and other actions and logs them. Activity associated with user accounts (where applicable). |

Appendix 2

Questions to the Ethical User Story Development Team

1. How did you experience this Ethical User Story activity?

2. How did you experience the process of creating this Ethical User Story activity?

3. Feedback on the activity:

# List of references: